\documentclass[pra,twocolumn,showpacs,superscriptaddress,10pt]{revtex4-2}
\usepackage{amsmath,amssymb,graphicx,color}
\usepackage{epstopdf}

\begin{document}

\title{Optimizing Single-Photon Quantum Radar Detection through Partially Postselected Filtering}

\author{Liangsheng Li}
\email{liliangshengbititp@163.com}
\affiliation{National Key Laboratory of Scattering and Radiation, Beijing 100854, China}

\author{Maoxin Liu}

 \affiliation{School of Systems Science  $\&$ Institute of Nonequilibrium Systems, Beijing Normal University, Beijing 100875, China}

\author{Wen-Long You}
\affiliation{College of Physics, Nanjing University of Aeronautics and Astronautics, Nanjing, 211106, China}

\author{Chengjie Zhang}
\affiliation{School of Physical Science and Technology, Ningbo University, Ningbo, 315211, China}

\author{ Shengli  Zhang}
\affiliation{Center for Quantum Technology Research, School of Physics, Beijing Institute of Technology, Beijing 100081, China}

\author{Hongcheng Yin }
\affiliation{National Key Laboratory of Scattering and Radiation, Beijing 100854, China}

\author{Zhihe Xiao}
\affiliation{National Key Laboratory of Scattering and Radiation, Beijing 100854, China}

\author{Yong  Zhu}
\affiliation{National Key Laboratory of Scattering and Radiation, Beijing 100854, China}

\begin{abstract}

In this study, we explore
an approach aimed at enhancing the transmission or reflection coefficients  of absorbing materials through the utilization of joint measurements of entangled photon states. On the one hand, through the implementation of photon catalysis in the reflected channel, we can effectively modify the state of the transmission channel, leading to a notable improvement in the transmission ratio. Similarly, this approach holds potential for significantly amplifying the reflection ratio of absorbing materials, which is useful for detecting cooperative targets. On the other hand,  employing statistical counting methods based on the technique of heralding on zero photons, we evaluate the influence of our reflection enhancement protocol for detecting noncooperative targets, which is validated through Monte Carlo simulations of a quantum radar setup affected by Gaussian white noise. Our results demonstrate a remarkable enhancement in the signal-to-noise ratio of imaging, albeit with an increase in mean-square error. These findings highlight the potential practical applications of our approach in the implementation of quantum radar.

\end{abstract}

\maketitle 

\section{INTRODUCTION}

Radar is a highly effective device that utilizes electromagnetic signals to detect and determine the range of unknown targets~\cite{RMPradar}. Over the past few decades, scientists have conducted extensive theoretical and experimental research to enhance the  performance of realistic radar systems. In all radar systems, a high signal-to-noise ratio (SNR) plays a key role that determines the sensitivity and efficiency in realistic target detection. To improve the SNR, the most straightforward approach is to employ  optical
or microwave photon detectors that possess  both high sensitivity and high detection efficiency. Nowadays, ultra-sensitive detectors that are capable of counting single photons or even single microwave photons have been developed, providing a substantial enhancement in accuracy compared to conventional radar systems. The quantum effects at the single-photon level are negligible, undoubtedly heralding a new era of quantum radar where techniques from quantum information science (QIS) are harnessed for precise positioning, ranging, and detection of conventional targets.
Meanwhile, QIS has become a flourishing interdisciplinary field that combines quantum mechanics,  computation and information theory. Using quantum mechanics principles, QIS adopts a revolutionary approach to encode, store, transmit, and manipulate information~\cite{Niesen,Horodecki}, which seeks to successfully surpass
the performance limits of classical information systems.
For example, quantum metrology, in which quantum entangled states are prepared as probes and interferometric measurements are conducted during the probe readout, can potentially reach the sensitivity limit dictated by Heisenberg's principle~\cite{Sciencequanmetro1, NPquanmetro1, PRLquanmetro1, PRLquanmetro2}. This implies that the number of repetitions needed to achieve a certain level of precision is merely the square root of that required in classical measurement strategies, providing a quadratic acceleration to the measurement process.

Entanglement quantum radar was pioneered by Lloyd~\cite{Sciencequanill} and is alternatively referred to as quantum illumination radar,
which represents a genuine quantum radar system, wherein quantum entanglement is sent to interrogate with the target and joint quantum measurements are used to detect the resulting echo signal.
Many sophisticated techniques, such as efficient entanglement generation, optimized discrimination of quantum states, manipulation of photon numbers using photon catalysis, and high-resolution single-photon detection  in optical and microwave regimes, can be seamlessly integrated into quantum radar systems, making them increasingly suitable for various commercial and military applications.
Various types of quantum states have undergone testing and evaluation in the context of entanglement quantum radar,
including photon number correlation state~\cite{Meda_2017}, $N$-entangled photons\cite{QRsmith}, two-mode squeezing state (TMSS)~\cite{QRluong}, photon-subtracted TMSS~\cite{PhysRevA.89.062309}, photon-added TMSS~\cite{qinongaussMSZ}, multi-mode quantum entanglement state~\cite{Jung2022}, X-ray quantum entanglement~\cite{PhysRevLett.127.013603} and even microwave quantum entanglement state~\cite{Barzanjeheabb0451}.
Typically, achieving a 6dB advantage over conventional radar through entanglement quantum illumination necessitates the implementation of joint quantum measurements on all echo signals and auxiliary entangled modes~\cite{PhysRevLett.101.253601}.
To alleviate the substantial demands on quantum storage capacity, more efficient joint-measurement-free measurement schemes become imperative.
A mode-by-mode scheme
for quantum receiver with balanced homodyne measurement
and optical parametric amplification was designed~\cite{diffCM}. S. Zhang {\it et. al.} proposed to use noiseless linear amplification to enhance receiver~\cite{Sheng-Lizhang}, which is also applicable to quantum illumination with Gaussian states~\cite{Karsa_2022}. An optimal receiver utilizing iterative sum-frequency generation was proposed, demonstrating the possibility of achieving the quantum Chernoff bound for quantum illumination in the low-signal brightness limit~\cite{PhysRevLett.118.040801}.

To further explore the unprecedent potential of the
quantum radar, various characteristics of quantum radar
have been theoretically studied.
The quantum radar cross-section (QRCS) was introduced via quantum electrodynamics and interferometric principles~\cite{QRlanzagorta2010} and has mainly been explored theoretically. A closed-form expression for and analysis of the slumping effect of a cuboid in the scattering characteristics of quantum radar were derived~\cite{Tian:21}. Furthermore, a method for calculating the orthogonal projected area of a target in each incidence, which is a key component in determining the QRCS of any arbitrary three-dimensional convex target~\cite{QRfang}.   Entanglement quantum radar can 
enhance various applications, including data reading~\cite{qread,Ortolanoeabc7796}, imaging~\cite{Gregoryeaay2652},  velocity measurement~\cite{PhysRevA.96.040304},  target ranging~\cite{PhysRevLett.128.010501} and biomedical imaging~\cite{Zheltikov_2020,Aslam2023}.
Additionally, research has explored the enhancement of 
LiDAR through quantum entanglement~\cite{QRlanzagorta2013}.
More importantly,  recent experiments in quantum radar have made 
significant progress~\cite{PRLqrexp1, APLqrexp1,QRmenzel}.

However, most of the results concerning entanglement quantum radar assume cooperative targets that are highly reflective and lossless.
In practical scenarios, it is more likely that the target is noncooperative or even offensive, 
 especially when covered with absorbing materials.
 In such cases, the reflectivity or transmissibility of the target is extremely low, posing a challenge in collecting sufficient echo signals for effective  quantum measurements.   Consequently,  a photon filtering strategy that can potentially improve the transmission or reflection coefficients of noncooperative targets becomes significantly important in all quantum radar systems.
 In this work, we propose a photon-filtering scheme to enhance the transmission coefficients, which can be easily adapted to enhance the reflection coefficients. Moreover, we introduce two methods to implement the photon filtering scheme. The first method is based on photon catalysis~\cite{PhysRevLett.88.250401,PhysRevA.68.032310,PhysRevA.92.012318,PhysRevA.97.043830,PhysRevA.99.032327,Tang2012QuantumWheeler}  
 and the second method involves the technique of heralding on zero photons~\cite{Nunn2021}. Both approaches can significantly improve the imaging of absorbing targets.

The rest of the paper is organized as follows. In Sec. \ref{sec_2},
we illustrate the effective enhancement of transmission or reflection signals through the utilization of a partially postselected quantum filter. Section \ref{sec25} introduces the implementation of the photon catalysis,  which is useful for detecting cooperative targets.  Based on the technique of heralding on zero photons \cite{Nunn2021}, we present another photon-filtering scheme for detecting noncooperative targets in Sec. \ref{sec_3}. We employ statistical counting methods to simulate the effects of reflection enhancement, and demonstrate the effectiveness of our enhancement scheme in significantly improving the SNR of imaging through a Monte Carlo simulation on Gaussian white noise.
Finally, Sec. \ref{sec_4} summarizes our results.

\section{Partially Postselected Filter}\label{sec_2}
In this section, we illustrate how the
transmission signal or the reflection signal
is effectively enhanced. The main idea is to manipulate a quantum state to be measured by encoding 
a prepared parameter.
To be concrete, the
photon filtering is employed to enhance the target signal.
 The photon filtering employed in this work utilizes a form of postselection~\cite{Arvidsson-Shukur2020,PhysRevLett.128.220504}.
Past research has shown that postselection can significantly alter photon statistics~\cite{gottesman1999demonstrating,bouwmeester1997experimental,Ajiki_2009,mahler2016experimental}. 
 Recent theoretical research has suggested that postselected quantum experiments have the potential to surpass the Heisenberg limit by allowing quantum states to carry additional Fisher information~\cite{PhysRevA.107.042413}. 
Such benefit may be connected to the negative quasiprobability
distribution~\cite{RevModPhys.17.195,PhysRev.44.31,PhysRevA.76.012119,PhysRevLett.101.020401,PTP.26.722}. The improved quantum advantages
can be attained when properly conditioned experiments are performed  with a lower
rate of successful postselection~\cite{PhysRevA.88.042116,qc9}.  Notably, in a recent polarimetry experiment, a quantum postselection protocol substantially elevated the precision of per-detected photons by more than two orders of magnitude~\cite{PhysRevLett.128.220504}. These progresses inspire us to explore the potential for improving the measurement of the SNR in reflections by devising a postselection protocol involving an absorbing material.

 Consider a
 photon state
\begin{equation}
|s\rangle = \sum_{n} c_n|n\rangle,
\end{equation}
where $|n\rangle$ represents a photonic Fock state. The photon filtering based noiseless quantum amplifiers have been produced using the nonunitary operation~\cite{PRAQS2022}
\begin{equation}
G_{s c}=\frac{1}{g}(|0\rangle\langle 0|+g| 1\rangle\langle 1|),\label{QSc}
\end{equation}
where $g>1$ is the gain.
Then applying
$G_{s c}$ acting on the state $|s\rangle$,
one readily
obtains a state
\begin{equation}
|qs\rangle = \frac{c_0}{g}|0\rangle+gc_1|1\rangle.
\end{equation}
The components other than $|0\rangle$ and  $|1\rangle$   are eliminated from the state $|qs\rangle$. 
For convenience,
we apply the partially postselected filter, which is defined by the operator via conditional measurements:
\begin{equation}
F=|0\rangle\langle 0|+p| 1\rangle\langle 1|,\label{Eq4}
\end{equation}
in which $p$ is a postselection parameter.
We will see in the
following a partially postselected quantum filter can be  
utilized to effectively improve both the transmission and the reflection of a single-photon state input.
  
\begin{figure}[tp]
	\centering
	\includegraphics[scale=0.18]{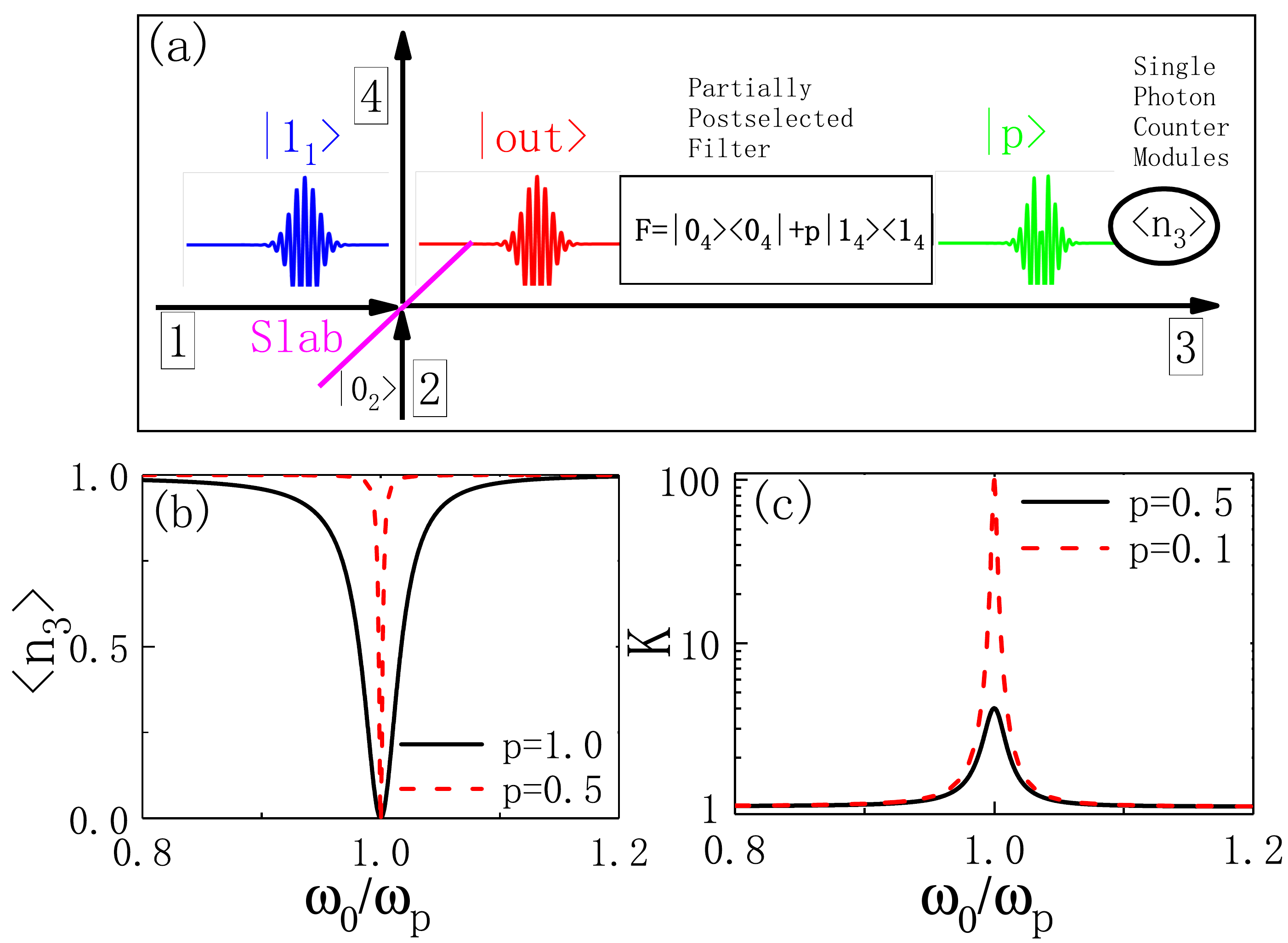}%
	\caption{ (a) An implementation of partially postselected filter via conditional measurements in a beam splitter. A single photon is prepared in the input mode of a dispersion beam splitter with reflectivity $\left| r(\omega ) \right|^{2}$. The incident angle is $\theta =\pi /4$.  The performance of partially postselected filter with different
postselection parameter. (b) The average photon number in channel 3 
versus incident frequency. (c) The amplification ratio 
versus incident frequency. Increasing $p$ reduces the amplification. }
\label{transmission}
\end{figure}

\textit{Enhanced transmission based on a quantum filter.}
First, we discuss the scheme
to enhance the transmission by the partially postselected quantum filter,
as is illustrated in Fig. \ref{transmission} (a).
One can find the setup consists of three parts, including
two-channel scattering, partially postselected quantum filter and a signal
counter module.
The two-channel scattering is implemented by a beam splitter, which is a finite thickness slab as shown in Fig. \ref{transmission} (a). There are two channels of inputs, including an input mode-1 of a single photon state and an input mode-2 of the
vacuum state,
which are divided
by the beam splitter into the outputs of mode-3 and mode-4.
Supposing that in the dispersive slab
the absorption might be neglected,
the input-output formalism between the annihilation operators of input modes 
 ${{\vec{P}}^{T}}={{\left( {{a}_{1}}(\omega ),{{a}_{2}}(\omega ) \right)}^{T}}$and output modes 
${{\vec{O}}^{T}}={{\left( {{a}_{3}}(\omega ),{{a}_{4}}(\omega ) \right)}^{T}}$ can then be described by a linear relation as
\begin{equation}
\vec{O}=\mathcal{T}(\omega) \cdot \vec{P},
\end{equation}
where the operators of each
channel are
assumed to satisfy the following commutation relations,
\begin{equation}
\begin{aligned}
& {\left[a_\alpha(\omega), a_\beta^{\dag}\left(\omega^{\prime}\right)\right]=\delta\left(\omega-\omega^{\prime}\right) \delta(\alpha-\beta)}, \\
& {\left[a_\alpha(\omega), a_\beta\left(\omega^{\prime}\right)\right]=\left[a_\alpha^{\dag}(\omega), a_\beta^{\dag}\left(\omega^{\prime}\right)\right]=0}.
\end{aligned}
\end{equation}
The transformation matrix $\mathcal{T}(\omega )\in \textrm{SU}(2)$ is 
a unitary matrix, i.e.,
 ${{\mathcal{T}}^{\dag}}(\omega )\mathcal{T}(\omega )=I$,  and 
 thus can be explicitly expressed by
\begin{equation}
\mathcal{T}(\omega)=\left(\begin{array}{cc}
t(\omega) & r(\omega) \\
-r^*(\omega) & t^*(\omega)
\end{array}\right).
\end{equation}
Here $r(\omega )$and $t(\omega )$
are reflection and transmission coefficients, respectively.

In order to
determine the transformation matrix $\mathcal{T}(\omega )$ of the dispersive slab, we
adopt the lossless Drude model, where the permittivity is $\varepsilon =1-\omega _{p}^{2}/\omega _{{}}^{2}$ and the
permeability is $\mu =1$.
Without loss of generality, we choose a set of typical parameters of the beam splitter. The plasma frequency is
$\omega _{p}^{{}}={{10}^{14}}$ Hz and the slab thickness is $L=3 \mu \text{m}$. Thus,
the reflection and transmission coefficients of beam splitter model with an incident angle  $\theta$ is given by
\begin{eqnarray}
t(\omega)=\frac{\alpha}{\beta^2-1} e^{-i \bar{\omega} \cos \theta},  \quad
r(\omega)=\frac{\alpha \beta-\beta^2+1}{\beta^2-1},
\end{eqnarray}
where the parameters
\begin{eqnarray}
\bar{\omega}&=&\frac{\omega}{\omega_p},\nonumber \\
\alpha&=&4 D \bar{\omega} \cos \theta\left(D^2+\sin ^2 \theta\right)\left(e^{-i D}-e^{i D}\right),
\nonumber \\
\beta&=&\frac{\left(D^2+\sin ^2 \theta\right)\left(e^{-i D}-e^{i D}\right)+D \bar{\omega} \cos \theta\left(e^{-i D}+e^{i D}\right)}{2 D \bar{\omega} \cos \theta}, \nonumber \\
 D&=&\sqrt{\left(\bar{\omega} \cos \theta\right)^2-1}.
\end{eqnarray}

With an explicit description of  transformation matrix $T(\omega )$, we can accurately compute the scattering process of the beam splitter.
To proceed, we focus on a single photon state residing in mode-1, characterized by solely positive frequencies,
which is defined as
\begin{equation}
\left|1_1\right\rangle=\int_0^{\infty} d \omega \Gamma(\omega) a_1^{\dag}(\omega)|0_1\rangle,
\end{equation}
where $\Gamma(\omega)$ is  the frequency distribution function.
The mode-2 is initially prepared in the vacuum state  $\left| {{0}_{2}} \right\rangle $.
Thus the input state is
\begin{equation}
| \mathrm{ in }\rangle=\int_0^{\infty} d \omega \Gamma(\omega) a_1^{\dag}(\omega)|0_1\rangle \otimes \left| {{0}_{2}} \right\rangle,
\end{equation}
satisfying
\begin{equation}
\langle\mathrm{in}| \mathrm{in}\rangle=\int_0^{\infty} d \omega|\Gamma(\omega)|^2=1.
\end{equation}

According to the input-output
formalism, the
output mode can be given by
\begin{equation}\label{outstate}
|\text {out}\rangle=\int_0^{\infty} d \omega \Gamma(\omega)\left[t(\omega) a_3^{\dag}(\omega)-r^*(\omega) a_4^{\dag}(\omega)\right]|0\rangle.
\end{equation}
The state $|\text {out}\rangle$ is an
entangled state. To clearly see it, we
rewrite Eq.(\ref{outstate}) as
\begin{eqnarray}\label{outstate_en}
|\text {out}\rangle=\int_0^{\infty} d \omega \Gamma(\omega)\left[t(\omega)|1^{\omega}_3\rangle \otimes|0^{\omega}_4\rangle-r^*(\omega)|0^{\omega}_3\rangle \otimes|1^{\omega}_4\rangle\right].~~
\end{eqnarray}
We then apply a partially postselected filter, which is given by
\begin{eqnarray}
F_T=|0_4\rangle\left\langle 0_4|+p| 1_4\right\rangle\left\langle 1_4\right|, \label{Eq15} 
\label{FT152}
\end{eqnarray}
where the postselection parameter $\left| p \right|\in \left[ 0,1 \right]$.
Here the partially postselected filter acts as a Kraus operator~\cite{PRAQS2022} for the single-photon state input. The state $F_T \vert \text {out}\rangle$ can be easily obtained by replacing $r^{*}$ with $pr^{*}$ in Eq.\eqref{outstate}, namely,
\begin{eqnarray}
F_T|\text {out}\rangle=  \int_0^{\infty} d \omega \Gamma(\omega)\left[t(\omega) a_3^{\dag}(\omega)-pr^*(\omega) a_4^{\dag}(\omega)\right]|0\rangle.~~
\end{eqnarray}

If we can perform a quantum state
filtering on the field in the mode-4, then the filter
allows $\left| \text{out} \right\rangle$ state to pass with a probability
\begin{eqnarray}
A=\left\langle\text {out}\left|F^{\dag}_T F_T\right| \text {out}\right\rangle.
\end{eqnarray}
We now focus on
the photon number in the partially postselected quantum detection experiments. The postselection is realized by a projective measurement
after a unitary evolution set by the beam splitter, 
while before the final photon detection. Then, the 
postselected state becomes
\begin{eqnarray}
|p\rangle=F_T |\text {out}\rangle / \sqrt{A}. 
\end{eqnarray}
Note that this state should be renormalized by the passing probability, which
stems from the projective measurement implemented by the partially postselected filter
with limited efficiency.

\begin{figure}[tp]
	\centering
	\includegraphics[scale=0.20]{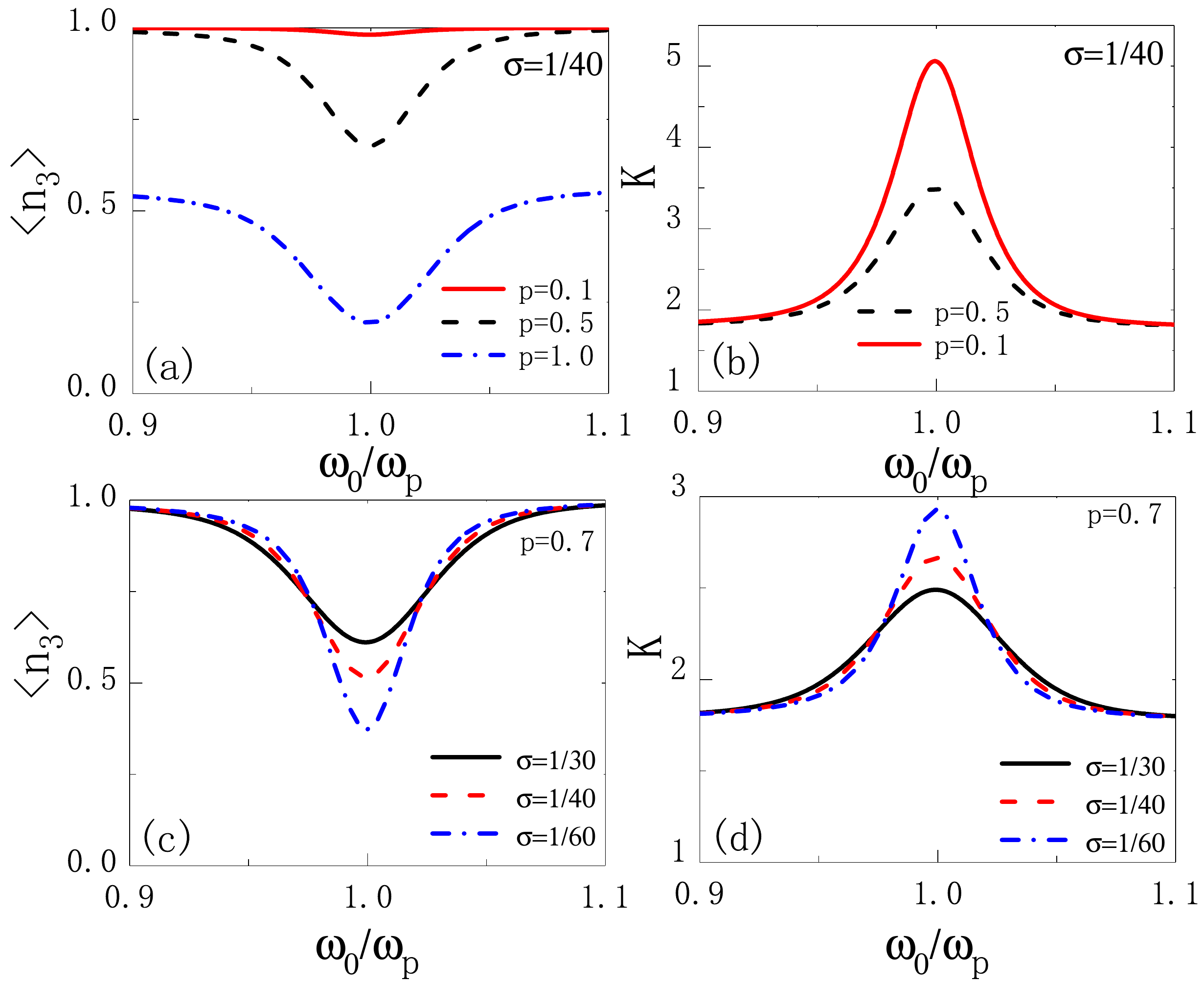}%
	\caption{ 
The results for Gaussian frequency distribution in Eq.~(\ref{Eq:Gaussiandistribution}). (a) The average photon number in channel 3 $\langle n_3\rangle $ and (b) the amplification ratio $K$ as a function of incident frequency with $\sigma =1/40$ and $p=0.1,~0.5,~1.0$. (c) The average photon number in channel 3 $\langle n_3\rangle $ and (d)  the amplification ratio  $K$ as a function of incident frequency with $\sigma =1/30,~1/40,~1/60$ and $p=0.7$.
  }
\label{Gaussian}
\end{figure}

However, the filter effectively amplifies the average photon numbers in the mode-3,
\begin{eqnarray}
\left\langle n_3\right\rangle_p=\frac{\int_0^{\infty} d \omega|\Gamma(\omega) t(\omega)|^2}{\int_0^{\infty} d \omega|\Gamma(\omega)|^2\left[|t(\omega)|^2+|p r(\omega)|^2\right]}.
\end{eqnarray}
To illustrate the amplification effect occurring specifically at the frequency $\omega_0$, we employ the Dirac delta distribution, i.e.,
\begin{eqnarray}
| \Gamma (\omega) |^2=\delta (\omega -\omega_0).\label{deltadistribution}
\end{eqnarray}
In this case, the mean photon number in channel-3 can be readily obtained by
\begin{eqnarray}
\label{n3}
\left\langle n_3\right\rangle_p=\frac{T\left(\omega_0\right)}{T\left(\omega_0\right)+|p|^2 R\left(\omega_0\right)}.
\end{eqnarray}
 One observes that the $|\text{out}\rangle$ state is recovered for $p=1$ with $ \langle n_3  \rangle _{p=1}=T(\omega _{0})<1$, which is equal to the transmission ratio
without the operation of the partially postselected filter. Another limit is the case of $p=0$ with $\langle n_3 \rangle _{p=0}=1$, which implies
that the effective transmission is enhanced. For a generic value of $0<p<1$,
one can find from Eq.(\ref{n3}) that  $\left\langle n_3\right\rangle_p$ is larger than $\left\langle n_3\right\rangle_{p=1}$, since positive valued $T\left(\omega_0\right)$ and $R\left(\omega_0\right)$ satisfy the constrain that $T\left(\omega_0\right)+ R\left(\omega_0\right)=1$. It clearly suggests that the partially postselected filter amplifies the average photon numbers in the channel 3. To better 
quantify the effect, we thus define  
an amplified ratio given by
\begin{eqnarray}
K=\frac{\left\langle n_3\right\rangle_p}{\left\langle n_3\right\rangle_{p=1}}.
\end{eqnarray}
The larger 
the ratio $K$ is, the better
the amplification effect is achieved.
A large average photon number is typically easier to be detected than a smaller one. If the average photon number ${{\left\langle {{n}_{3}} \right\rangle }_{p=1}}\ll 1$, then the partially postselected filter boosts the amplified ratio $K\approx \frac{1}{{{\left| p \right|}^{2}}}$. The amplified ratio 
can be tuned to be arbitrarily large if ${{\left\langle {{n}_{3}} \right\rangle }_{p=1}}$ is arbitrary small.

The enhancement scheme of transmission is applicable not only to the 
$\delta$-distribution in Eq.(\ref{deltadistribution}), but also to the general frequency distribution of the incident state. 
To illustrate this, we consider a Gaussian frequency distribution:
\begin{eqnarray}
\Gamma(\omega)=\frac{1}{\sqrt{\sigma \pi}}e^{-i \frac{\omega-\omega_0}{ \omega_0 \tau}} e^{-\frac{1}{2 \sigma^2}\left(\frac{\omega-\omega_0}{\omega_0}\right)^2}.
\label{Eq:Gaussiandistribution}
\end{eqnarray}
Figure \ref{Gaussian} (a) illustrates how the average transmitted photon number $\langle n_3\rangle$ is affected by varying the value of $p$, while keeping the width of the photon field frequency distribution fixed. As the value of $p$ decreases, the efficiency of the transmission is improved. This effect is more prominently demonstrated in Fig. \ref{Gaussian} (b), where the gain $K$ increases significantly with decreasing values of $p$.
Figures \ref{Gaussian} (c) and \ref{Gaussian} (d) depict the impact of the width of the frequency distribution of the photon field on the amplification of transmission in the amplification scheme. It is noteworthy that a narrower frequency distribution enhances the amplification effect, particularly in the vicinity of the center of the distribution denoted as $\omega_0$.
\section{Implementation of photon filtering by photon catalysis}\label{sec25} 
\begin{figure}[htpb]
	\centering
	\includegraphics[scale=0.25]{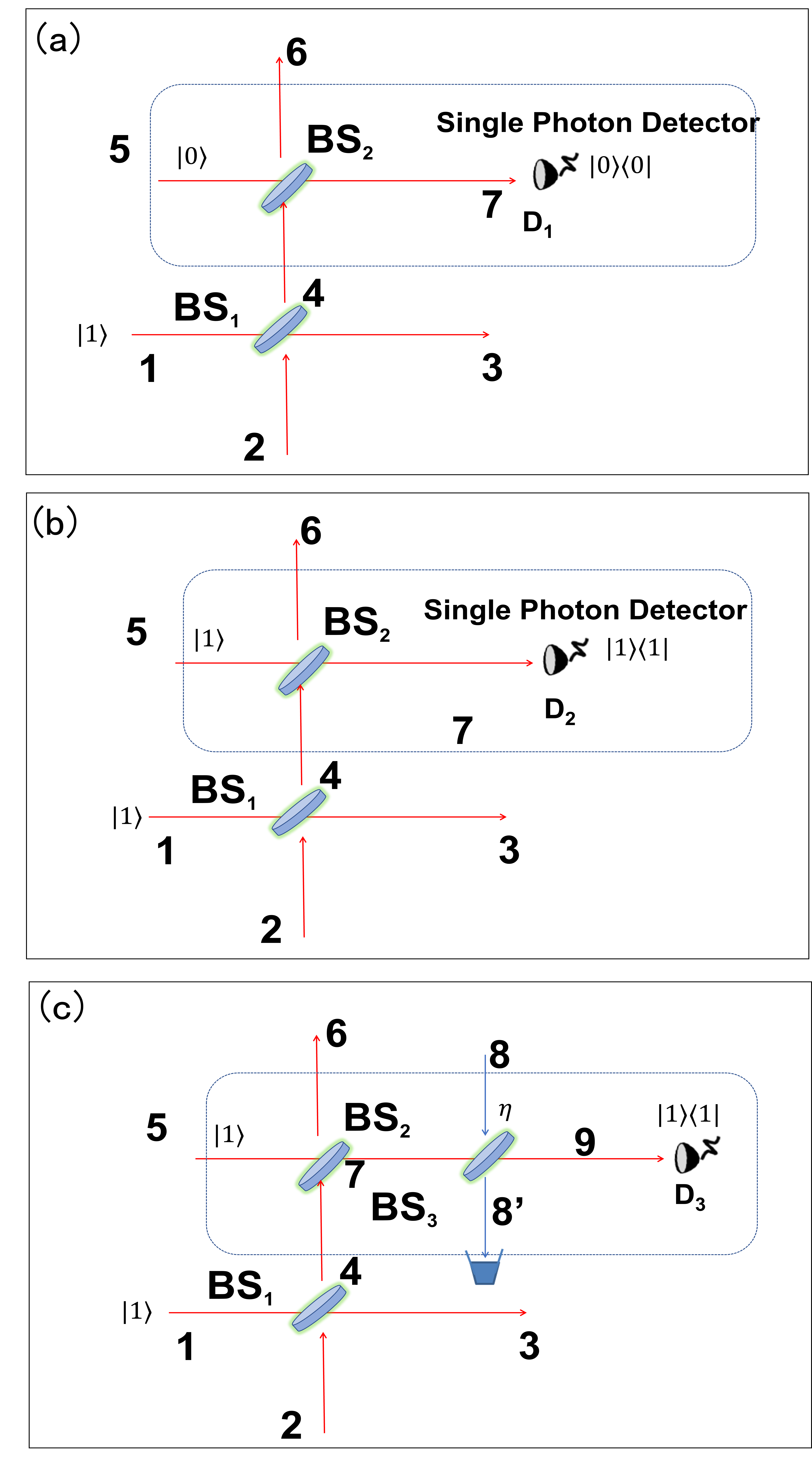}%
	\caption{ A sketch for implementing quantum catalysis using  (a) zero-photon detection, (b) ideal single-photon detection, and (c) realistic single-photon detection. All detector D$_1$, D$_2$ are ideal single-photon detectors. The beam splitter BS$_3$ with transmittance of $\eta$, combined with the ideal single-photon  detector D$_3$ capable of photon number resolution, is used to simulate a realistic single photon detector with detection efficiency of $\eta$. The input optical quantum states in mode 2 and 8 are vacuum state $|0\rangle$.}
\label{qs}
\end{figure}

Section \ref{sec_2} introduces a general approach for enhancing transmission or reflection channels via photon filtering. However, the specific implementations of the photon filtering operation in Eq. (\ref{FT152}) have not been discussed yet. To address this gap, we will present two methods for implementing Eq. (\ref{FT152}). The first method, based on photon catalysis technology, is primarily utilized for cooperative targets.
Figure \ref{qs} shows implementations of the photon catalysis, which effectively implements the filter in Eqs. (\ref{Eq4}) and (\ref{Eq15}). In Fig. \ref{qs} (a), the transmission ratio
of beam splitter BS$_1$ is $T$ and BS$_2$ is $p^2$.
The input in channel-1 is a single-photon state, while channels 2 and 5 are in a vacuum state. If the single photon detector obtains zero photon, the entangled state of channel-3 and channel-6 becomes
\begin{eqnarray}
|\psi\rangle=\frac{\sqrt{T}|1\rangle|0\rangle+p \sqrt{1-T}|0\rangle|1\rangle}{\sqrt{T+(1-T) p^2}},
\end{eqnarray}
from which we obtain the measured photon number in channel-3 as follows,
\begin{eqnarray}
\langle n\rangle_p=\frac{T}{T+p^2(1-T)},
\end{eqnarray}
which recovers Eq. \eqref{n3}.

However, the scheme in  Fig. \ref{qs} (a) needs a single photon detector with $100\%$ accuracy, or the single photon state  $|1\rangle$ would be faultily considered as a vacuum $|0\rangle$. 
To accurately detect the zero photon state at detector $D_1$, as shown in  Fig.\ref{qs} (a), a heralding technique on detection of zero photons is necessary~\cite{Nunn2021}. This, however, requires the addition of a pulsed laser light source, increasing the complexity and cost of the setup. In this study, we propose alternative methodologies that circumvent the need for detecting the zero photon state. Specifically,  
 we design an alternative scheme as shown in Fig. \ref{qs} (b). 
 If the single photon detector $D_2$ obtains a single photon ($D_2$ is a photon-number-resolving detector, which can distinguish 1 from 2 photons), the entangled state of channel-3 and channel-6 becomes
\begin{eqnarray}
|\psi\rangle=\frac{p \sqrt{T}|1\rangle|0\rangle+\left(2 p^2-1\right) \sqrt{1-T}|0\rangle|1\rangle}{\sqrt{p^2 T+(1-T)\left(2 p^2-1\right)^2}},\label{psip}
\end{eqnarray}
from which we obtain the measured photon number in channel-3 as follows,
\begin{eqnarray}
\langle n\rangle_p=\frac{T p^2}{p^2 T+(1-T)\left(2 p^2-1\right)^2},
\end{eqnarray}
which also enhances the photon number.

\begin{figure}[htpb]
	\centering
	\includegraphics[scale=0.5]{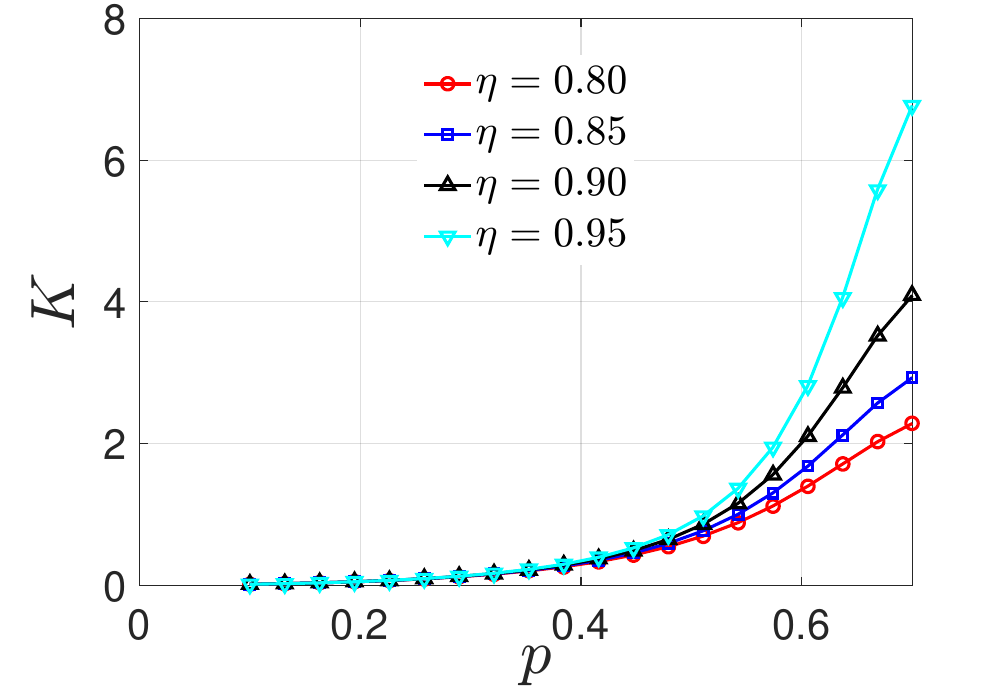}%
	\caption{ Amplification ratio  $K$ as a function of $p$. $\eta$ is chosen to be $0.80,0.85,0.90,0.95$, from bottom to top.}
\label{KvalueWitheta}
\end{figure}

The scheme in Fig.\ref{qs} (b) also works when the detector is a realistic single-photon detector with non-unit detection efficiency. In Fig.\ref{qs} (c) we use a beam splitter BS$_3$ with
transmissivity $\eta$ and an ideal single-photon detector to represent a detector with efficiency $\eta$. The input mode of channel-2 and channel-8 are injected with the vacuum state
$|0\rangle $, and the outgoing mode channel-8$^\prime$ is discarded. When the detector D$_3$ registers a single photon, the photon catalysis process is considered successful. The probability of photon catalysis follows
\begin{eqnarray}
P_\mathrm{p,\eta}=\eta\left[p^2T +(1-T)(4p^4\eta-4p^2\eta+1)\right],
\end{eqnarray}
and the output state in channel 3 and 6 are now being a mixed state given by
\begin{eqnarray}
\rho_{p,\eta}=\frac{4p^2(1-p^2)(1-T)\eta(1-\eta)|0\rangle|0\rangle\langle 0|\langle0|+|\widetilde{\psi}\rangle\langle \widetilde{\psi}|}{P_\mathrm{p,\eta}}  ,\quad
\end{eqnarray}
with $|\widetilde{\psi}\rangle$ being an unnormalized state
\begin{eqnarray}
|\widetilde{\psi}\rangle=\sqrt{\eta}[p\sqrt{T}|1\rangle|0\rangle+\sqrt{1-T}(2p^2-1)|0\rangle|1\rangle].
\end{eqnarray}
For $\eta\rightarrow 1$, it can be easily checked that $|\widetilde{\psi}\rangle$ (if normalized) is exactly the state in Eq.(\ref{psip}). Moreover, one can see that the non-unit detection efficiency induces a non-negligible vacuum state component  $|0\rangle|0\rangle$ in $\rho_{p,\eta}$, which does not ultimately contribute to the average photon number in channel-3. This explains why the scheme shown in  Fig.\ref{qs} (b) remains  robust even with a realistic single-photon detector. The average photon number in Fig.\ref{qs} (c) is now calculated as follows:
\begin{eqnarray}
\langle n\rangle_{p,\eta}=\frac{Tp^2}{p^2T +(1-T)(4p^4\eta-4p^2\eta+1) }
\end{eqnarray}
The amplification ratio in the case of the non-unit detector is given by 
\begin{eqnarray}
K_{p,\eta}=\frac{p^2}{p^2T +(1-T)(4p^4\eta-4p^2\eta+1)}.
\end{eqnarray}
A big amplification ratio, $K\gg 1$, can be observed even when using realistic detectors. In Fig.\ref{KvalueWitheta}, we show the amplification ratio for  $\eta$ values of $0.80,0.85,0.90,0.95$, respectively. For $p > 0.5$, a rapid increase of $K$ is observed.

\begin{figure}[tp]
	\centering
	\includegraphics[scale=0.5]{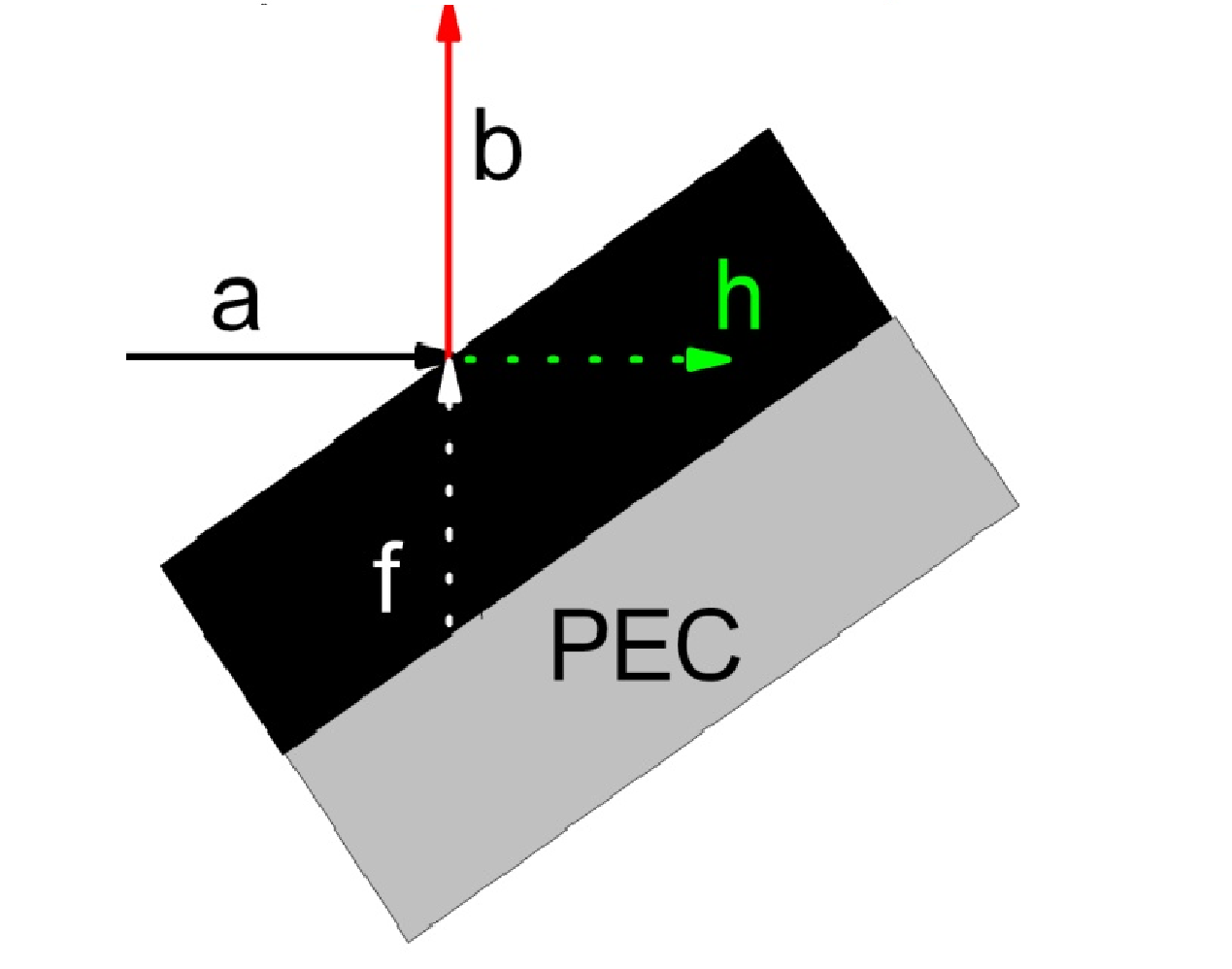}%
	\caption{A general schematic representation of a reflection system. "PEC" refers to a perfect electric conductor with total reflection, and the black portion represents its absorbing material surface. The input channel is labeled as "$a$", the reflection is denoted as "$b$," the noise from the absorbing surface is represented as "$f$" and the transmission channel is labeled as "$h$". }
\label{reflect}
\end{figure}

\textit{Quantum Radar aiming to detect the absorption.}
In an analogy way, the above scheme to enhance the transmission can be applied to improve the detection of the reflected signal and thus can be employed to build a quantum radar  for detecting cooperative targets. Figure  \ref{reflect} depicts
a reflection process on an absorbing material.  The incidents received through channel-$a$ are split into a reflected channel and an absorbed channel on the surface, known as channel-$b$ and channel-$h$. A noise caused by the material of channel-$f$
should be included.
Comparing the transmission process in Fig.~\ref{transmission}, there is a simple mapping relation between the indices of channels $\{1,2,3,4\}$ and $\{a,f, h,b\}$.

\begin{figure}[tp]
	\centering
	\includegraphics[scale=0.20]{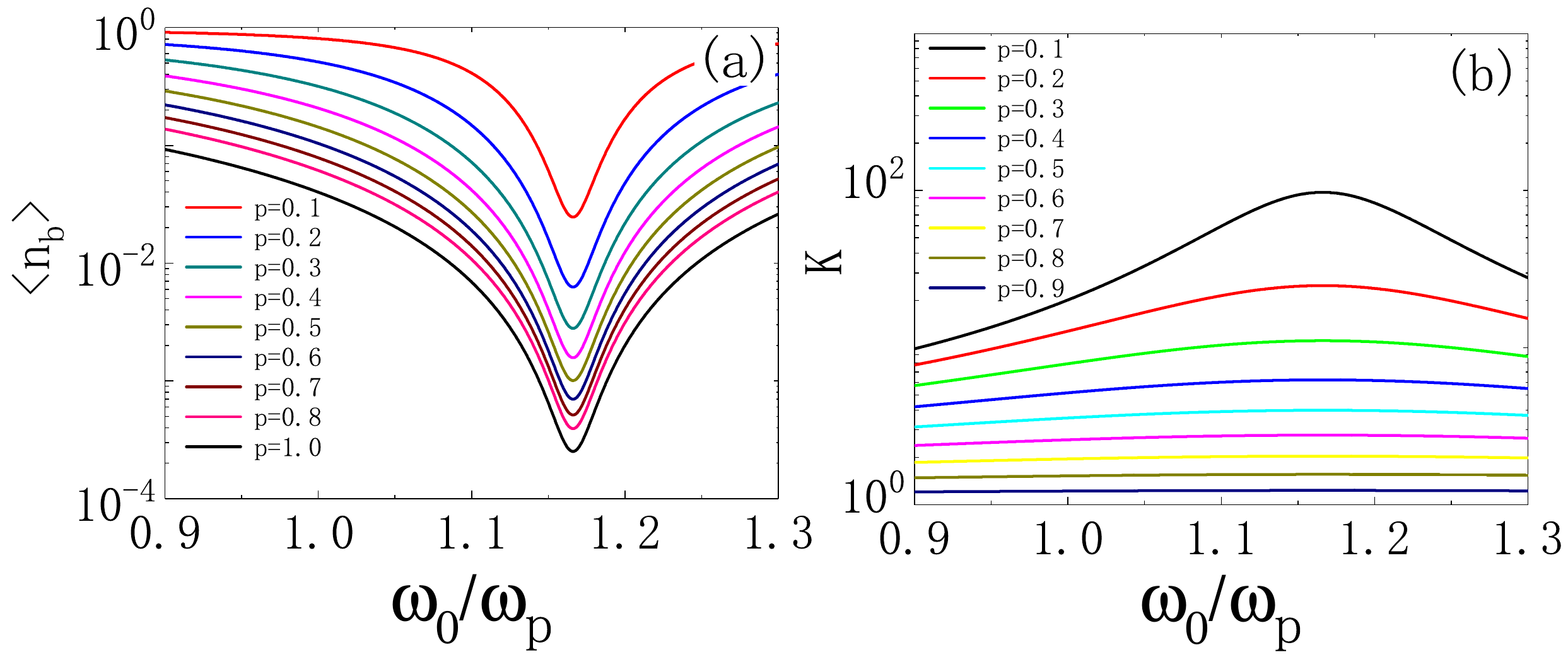}%
	\caption{
 (a) The average photon number in channel $b$ and (b) the corresponding amplification ratio $K$ as a function of incident frequency with different values of $p$  under $\delta$-distribution.
 }
\label{reflectenhance}
\end{figure}

The general relation between the input and output operators including loss is
\begin{eqnarray}
\left(\begin{array}{l}
b(\omega) \\
h(\omega)
\end{array}\right)=\mathcal{T}(\omega) \cdot\left(\begin{array}{l}
a(\omega) \\
f(\omega)
\end{array}\right),
\end{eqnarray}
where
\begin{eqnarray}
\mathcal{T}(\omega)=\left(\begin{array}{cc}
r(\omega) & k(\omega) \\
-k^*(\omega) & r^*(\omega)
\end{array}\right),
\end{eqnarray}
with
\begin{eqnarray}
|r(\omega)|^2+|k(\omega)|^2=1.
\end{eqnarray}
Thus, the input operators and output operators obey the commutation relations, i.e.,
$\left[\eta(\omega_1), \eta^{\dag}(\omega_2)\right]=\delta(\omega_1-\omega_2)$,
$\eta = a, b, h, f$, while they commute otherwise.
The input multimode single photon state at channel $a$ can be expressed as
\begin{eqnarray}
 |\mathrm{in}\rangle=\int_0^{\infty} d \omega \Gamma(\omega) a^{\dag}(\omega) \mid 0\rangle.
\end{eqnarray}
The scattered output field should be
\begin{eqnarray}
|\mathrm{out}\rangle=\int_0^{\infty} d \omega \Gamma(\omega)\left[r(\omega) b^{\dag}(\omega)-k^*(\omega) h^{\dag}(\omega)\right]|0\rangle.
\end{eqnarray}
Consider a frequency independent quantum filter
\begin{eqnarray}
F_R=\left|0_h\right\rangle\left\langle 0_h|+p| 1_h^\omega\right\rangle\left\langle 1_h^\omega\right|. 
\end{eqnarray}
Under the partially postselected filter, the
average photon number of multimode single photon scattering state 
in channel $b$ is
\begin{eqnarray}
\left\langle n_b\right\rangle=\frac{\int_0^{\infty} d \omega|\Gamma(\omega) r(\omega)|^2}{|p|^2+\left(1-|p|^2\right) \int_0^{\infty} d \omega|\Gamma(\omega) r(\omega)|^2},
\end{eqnarray}
and the amplified ratio is
\begin{eqnarray}
K=\frac{1}{|p|^2+\left(1-|p|^2\right) \int_0^{\infty} d \omega|\Gamma(\omega) r(\omega)|^2}.
\end{eqnarray}
Figure \ref{reflectenhance} displays the relation between the parameter $p$ and the average photon number in channel $b$ under the $\delta$-distribution. In particular, Fig. \ref{reflectenhance} (a) highlights that as the value of $p$ decreases, the average photon number $\langle n_b\rangle $ gradually increases, confirming the effectiveness of our scheme in enhancing reflection. Furthermore, Fig. \ref{reflectenhance} (b) presents a visually striking representation, illustrating that the enhancement effect becomes more pronounced as $p$ increases.
\section{Implementation of photon filtering by heralding on zero photons}\label{sec_3}
A crucial ingredient in the
transmission-enhancement and reflection-enhancement schemes described above 
is the realization of the photon filtering scheme,
which implements operators $F_T$ and $F_R$, respectively.
To implement the photon filtering scheme as outlined in Eq. (\ref{FT152}), we propose a second method for detecting noncooperative targets. This method utilizes a heralding technique on zero photons~\cite{Nunn2021} to achieve signal enhancement. 

\begin{figure}[tp]
	\centering
	\includegraphics[scale=0.16]{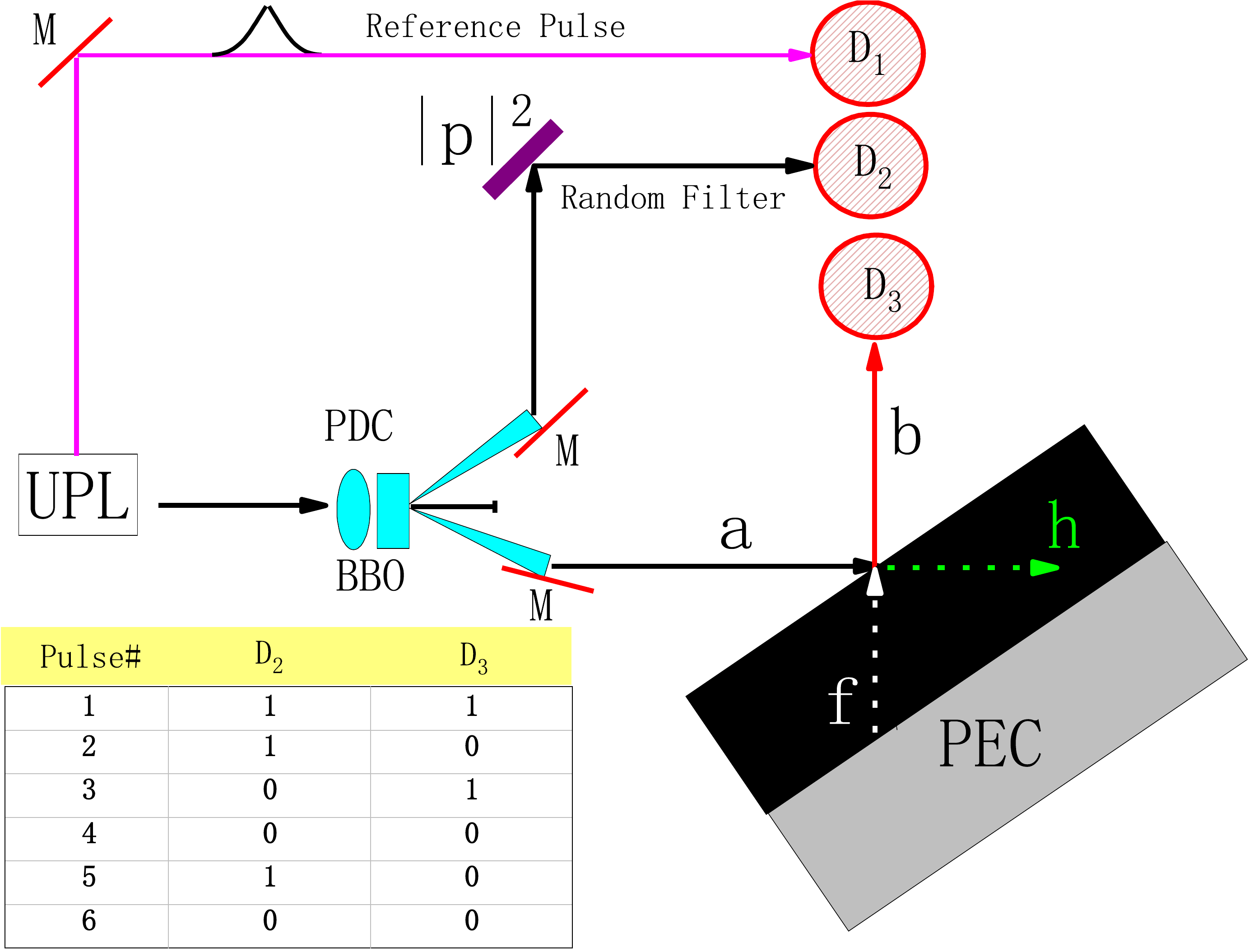}%
	\caption{An implementation of partially postselected filter by the technique of heralding on zero photons.
The lower left corner of the illustration showcases a list of correlated measurement data.} \label{classical}
\end{figure}

Figure \ref{classical} shows the
scheme we designed.
The basic physical process of this device is as follows.
A photon with a frequency of $2\omega$ is generated by the ultrapulse laser (UPL), and is subsequently converted into a pair of entangled photon both with a frequency of $\omega$ through a parametric down-conversion process using the $\beta$-barium borate crystal (BBO).
Two entangled photons enter the reflection system and auxiliary optical path, respectively. In the reflection system, the reflection and transmission 
of photons are determined by the reflection matrix. In other words, this photon has a probability of $\vert r \vert^2$ being captured by $D_3$ detector. In the auxiliary light path, another photon is reflected with a probability of $p^2$, and the probability of detection by the $D_2$  detector is also $p^2$.

In order to achieve signal enhancement, it is necessary to repeat the above process many times. Through this iterative procedure, we obtain a data list as depicted in the inserted illustration within Fig. \ref{classical}.
Suppose the number of entangled photon pairs produced by BBO is $N$. The $D_2$ and $D_3$ detectors are being measured at the same time.
 $N(D_2,D_3)$ denotes the times of coincidence counting that $D_2$ and $D_3$ simultaneously obtain readings. $N(D_2,0)$ indicates the number of times $D_2$ has a reading but $D_3$ has no reading. $N(0,D_3)$ indicates the number of times $D_3$ has a reading but $D_2$ has no reading. $N(0,0)$ indicates the number of times neither $D_2$ nor $D_3$ have readings at the same time. It is worth noting that $N(D_2,0)$, $N(0,D_3)$ and  $N(0,0)$ can be measured using the  technique of heralding on zero photons~\cite{Nunn2021}. 
 Based on the principle of probability, we can establish a relationship between $\{N,~N(D_2,D_3),~N(D_2,0),~N(0,D_3),~N(0,0)\}$ and $\{r,~p\}$. These entities are interconnected in the following manner:
\begin{eqnarray}
&&N(D_2,D_3)/N =R |p|^2, \nonumber \\
&&N(D_2,0)/N =  (1-R)|p|^2,\nonumber \\
&&N(0,D_3)/N = R(1-|p|^2),  \nonumber \\
&&N(0,0)/N =(1-R)(1-|p|^2).
\end{eqnarray}
Accordingly, we can construct a
measured quantity, given by
\begin{eqnarray}\label{yinyongchenggong}
M(T,p)
 \equiv \frac{N(1,1)+N(0,1)}{N-N(0,0)} =\frac{|r(\omega)|^2}{|p|^2+\left(1-|p|^2\right)|r(\omega)|^2}. \quad
\end{eqnarray}
Considering the relation that
\begin{eqnarray}
\left\langle n_b\right\rangle=\frac{|r(\omega)|^2}{|p|^2+\left(1-|p|^2\right)|r(\omega)|^2},
\end{eqnarray}\label{mtp}
we can readily find 
\begin{eqnarray}
M(T,p)=\left\langle n_b\right\rangle.
\end{eqnarray}
It means that the measurement of  $\left\langle n_b\right\rangle$ can be 
replaced by the statistical measurement
of $M$.

\begin{figure}[tp]
	\centering
	\includegraphics[scale=0.2]{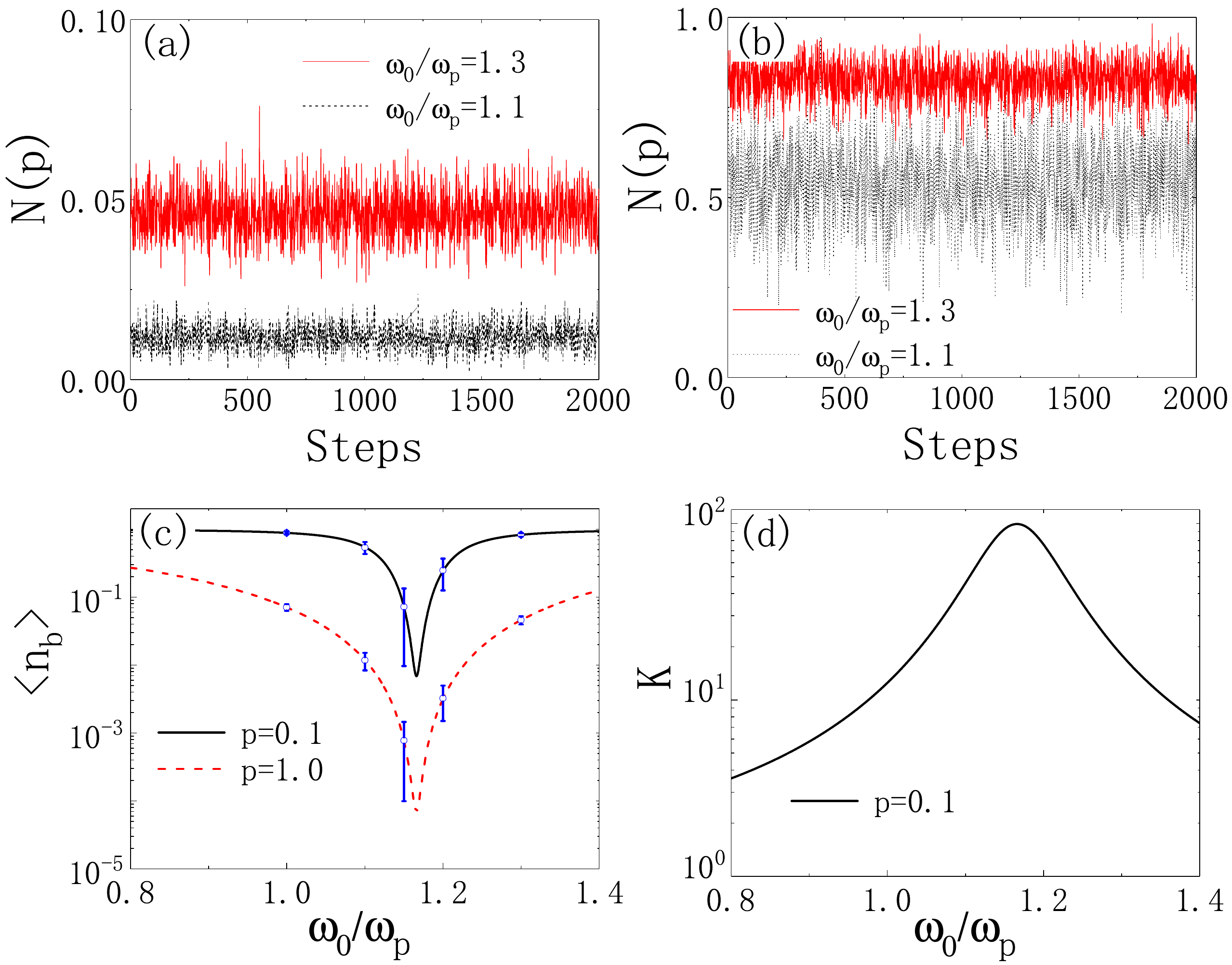}%
	\caption{ Illustration on details of the Monte Carlo simulation.  (a) and (b) show the simulation values of $N$ ($M(T,p)$) spanning 2000 Monte Carlo steps at different frequencies $\omega_0/\omega_p$ for two different values of $p$. Here $p=1$ for (a) and $p=0.1$ for (b), respectively. (c) The average effective reflection photon number as a function of frequency $\omega_0/\omega_p$ for different values of parameter $p$ ($1$ and $0.1$). (d) The gain $K$ as a function of frequency $\omega_0/\omega_p$ for a specific value of $p=0.1$.} \label{MonteCarlo}
\end{figure}

\begin{figure}[tp]
	\centering
	\includegraphics[scale=0.55]{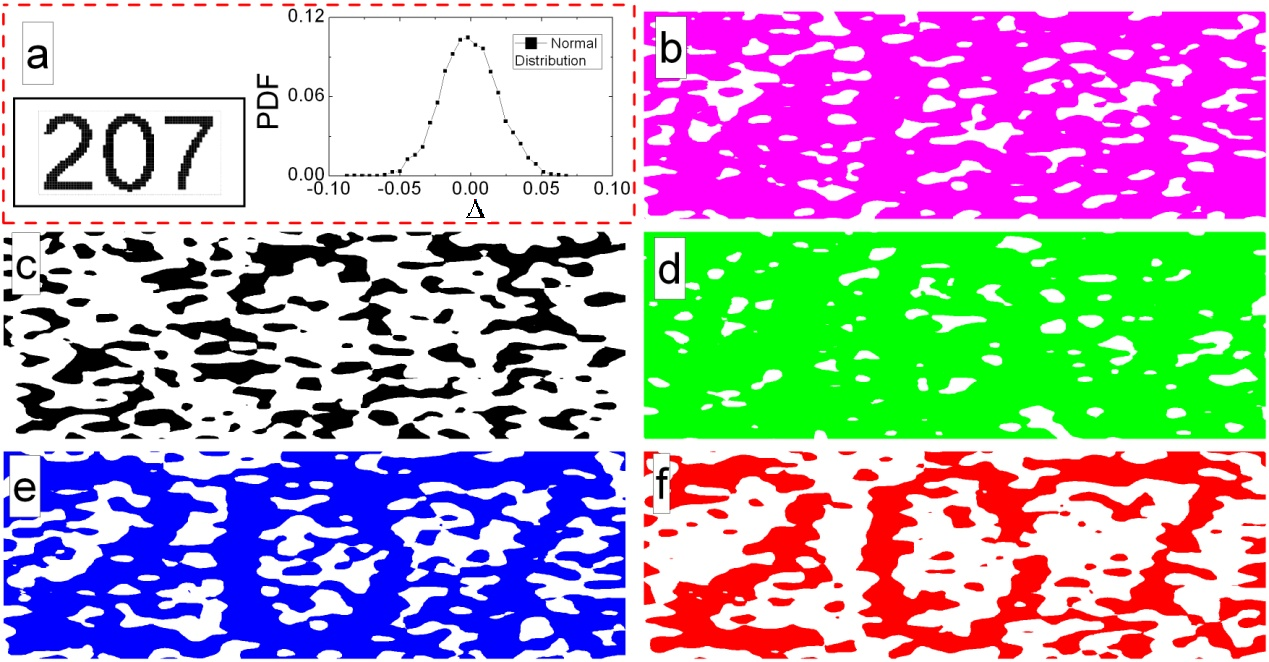}%
	\caption{ Illustration of imaging simulation. (a) The material with the pattern of the number "207", where the white part has the property of complete reflection, and the black part is made of absorbing material. The inset in (a) shows a normal noise distribution with the standard deviation $\sigma=0.02$. (b) to (f) depict the imaging results as the parameter $p$ gradually decreases with $p= 1.0, 0.8, 0.6, 0.4, 0.2$, respectively. }\label{207}
\end{figure}

In practical experiments, the value of $M(T,p)$ is not only determined by Eq.~\eqref{yinyongchenggong} but also includes fluctuations. Specifically, after considering the fluctuation effect,  $M(T,p)$ can be represented as
\begin{eqnarray}
M(T,p)=\frac{N(1,1)+\Delta N(1,1)+N(0,1)+\Delta N(0,1)}{N-(N(0,0)+\Delta N(0,0))}. \nonumber\\
\end{eqnarray}
Without loss of generality, 
$M(T,p)$ can be further simplified into
\begin{eqnarray}\label{mtpd}
M(T,p)=\frac{N(1,1)+N(0,1)}{N-N(0,0)} + \Delta.
\end{eqnarray}
The $\Delta$ term in Eq.\eqref{mtpd} encompasses all fluctuation effects and is assumed to follow a Gaussian noise distribution, characterized by
the probability density function (PDF) as
\begin{eqnarray}
f(\Delta)=\frac{1}{\sqrt{2\pi}\sigma}\exp(-\frac{\Delta^2}{2\sigma^2}).
\end{eqnarray}
The Monte Carlo (MC) simulation results of the $M(T,p)$ behavior are shown in 
Fig.~\ref{MonteCarlo}. The following outlines the steps of the MC simulation. First, given a frequency $\omega_0/\omega_p$ and a specific value 
of $p$, we generate a data list as depicted in 
Fig.~\ref{classical} under these conditions. In each row, the value of $D_2$ takes $1$ with probability $p^2$ and $0$ with probability $1-p^2$, and the value of $D_3$ takes $1$ with probability $\vert r \vert^2$ and $0$ with probability $1-\vert r \vert^2$. In our simulation process, the list length is 1000. Then, from the data list we can count four values of $N(D_2,D_3)$. In addition, we also need to consider the noise term in the formula in Eq.~(\ref{mtpd}), which is also calculated by random number simulation. Gathering all these, we can obtain the value of $M(T,p)$. In order to account for the average impact of various noise sources, we evaluate 2000 different noise samples for a given frequency  $\omega_0/\omega_p$ and a postselection parameter $p$. 
The simulation results for several specific frequencies  $\omega_0/\omega_p$  and $p$ parameters are shown in 
 Figs.~\ref{MonteCarlo} (a) and \ref{MonteCarlo} (b). By averaging these simulation results, we can obtain the relationship between the effective average reflected photon number $\langle n_b\rangle$ and frequency $\omega_0/\omega_p$. Figure \ref{MonteCarlo} (c) shows that 
for different $p$ values, $\langle n_b\rangle$ is a function of frequency $\omega_0/\omega_p$. It can be observed that when $p=0.1$, it effectively enhances $\langle n_b\rangle$ compared to $p=1$ (no optimization case) throughout the frequency domain. This enhancement effect is particularly significant near the resonance frequency. Figure \ref{MonteCarlo} (d) depicts the change in amplification rate with respect to the frequency.

Figure \ref{207} shows the imaging results of the medium material labeled with the number "207".
We depict imaging results as the parameter $p$ gradually decreases with $p= 1, 0.8, 0.6, 0.4, 0.2$, respectively.
In Fig.~\ref{207} (b), the original imaging result is shown with $p=1$, representing no enhancement effect. Due to noise, the image of the medium material is nearly indiscernible in this case. However, as the value of $p$ decreases, the enhancement effect gradually intensifies. Consequently, in Figs.~\ref{207} (c) to \ref{207}(f), the number "207" becomes progressively clearer. This visual demonstration serves as a clear and intuitive depiction of the substantial improvement in imaging enhancement achieved in this work.

\begin{figure}[tp]
	\centering
	\includegraphics[scale=0.37]{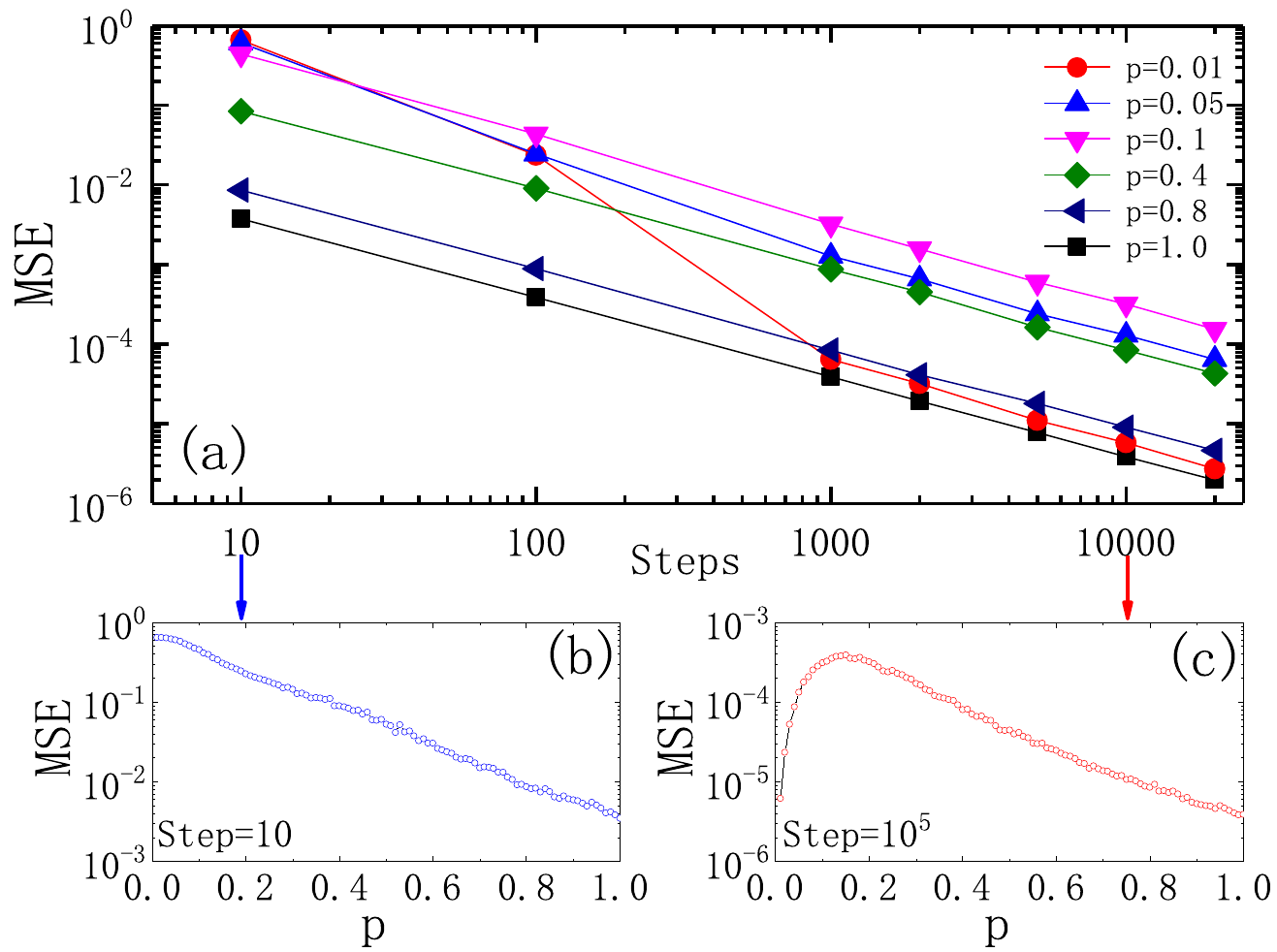}%
	\caption{ (a) The MSE versus the number of independent measurements for different values of the postselection parameter $p$ with $r$=0.2.
 The MSE is calculated by repeating  $N=1000$ photons in each measurement.  The MSE with respect to $p$ after performing a total of (b)  $\mathcal{N}_{\text{step}}=10$ and (c) $\mathcal{N}_{\text{step}}=10^4$ independent measurements.
	} \label{mse}
\end{figure}

Note that the photon filtering operation defined in Eq. (\ref{QSc}) can be implemented using the setup illustrated in Fig. \ref{classical}. Consequently, the application of the partially postselection method is justified. Furthermore,  it is essential to address the impact of data loss following the implementation of our scheme. Typically, the effectiveness of an estimation procedure is evaluated using the 
mean squared error (MSE)~\cite{PhysRevA.107.042413}, given by
\begin{equation}
\text{MSE}(p) = \frac{1}{\mathcal{N}_{\text{step}}} \sum_i  \left[M_i(T,p) - \langle n_b \rangle_i) \right]^2,
\end{equation}
in which $\mathcal{N}_{\text{step}}$ represents  the number of independent measurements.
Figure \ref{mse} shows the MSE for different step sizes with various values of $p$. In Fig. \ref{mse} (a), it becomes evident that the MSE for all $p < 1$ (when employing the enhanced scheme) exceeds that for $p = 1$ (without utilizing the enhanced scheme).
One observes that in Fig. \ref{mse} (b) the MSE exhibits an downward trend as $p$ increases. Intriguingly, at a step value of ${\mathcal{N}_{\text{step}}}=10^4$, we observe that although the MSE initially rises with decreasing $p$, it begins to decrease after surpassing a certain threshold, approximately around $p=0.15$.
Thus it is important to find a delicate balance between the improvement of the imaging SNR and the acceptable value of MSE for future practical applications, which has been successfully demonstrated in an optical experiment, such as partially postselected amplification~\cite{PhysRevLett.128.220504}.

\section{conclusion}\label{sec_4}
In this work, we  propose a scheme that enhances the transmission and reflection of absorbing materials. We achieve this by applying joint measurement to a pair of entangled photons,
with one used as a signal and the other as an auxiliary to enhance the signal. We first illustrate how our scheme can enhance the transmission of the incident photon state. The surface of the material acts as a beam splitter, dividing the incident photon state into the reflected and transmitted channels. Since the superposition state of these channels is entangled, we can use photon catalysis on the reflected channel to change the transmission channel state. By encoding the parameter $p$ into the transmissive state, we
show that the transmission is effectively enhanced.
Similarly, we can apply
the scenario to enhance the reflection of absorbing materials. The signal channel, vacuum channel, reflection channel, and transmission channel in the transmission enhancement scheme correspond respectively to the signal channel, noise channel, absorption channel, and reflection channel in the reflection enhancement scheme. We apply the photon catalysis to the absorption channels, and the parameter $p$ is encoded into the changed reflected state to enhance the reflected measurement. This results in an overall enhancement effect.

In order to experimentally implement reflection enhancement according to our scheme, we have devised an approach based on heralding on zero photons that applies statistical counting methods to simulate the effects of this enhancement. Specifically, we 
execute a Monte Carlo simulation on Gaussian white noise to validate our idea. 
The numerical tests have shown that our enhancement scheme is effective 
for significantly improving the imaging signal-to-noise ratio. The potential implications of our research are exciting, as our approach could have practical applications in quantum radar implementation.

\section*{Acknowledgments}
This work is supported by Ministry of Science and Technology of China (Grant No. 2016YFA0301300), National Science Foundation of China (Grants No. 11604009, 11474211, 12174194). We also acknowledge the computational support in Beijing Computational Science Research Center (CSRC).

\bibliography{ref_all}

\end{document}